\documentclass[dvips]{article}
\usepackage{supertabular,lscape,epsfig}
\usepackage{amssymb}
\usepackage{amsmath}

\DeclareSymbolFont{ppa}{OT1}{ppl}{m}{it}
\DeclareMathSymbol{\vv}{\mathalpha}{ppa}{'166}

\begin{document}

\newcommand{\dd}{\,{\rm d}}
\newcommand{\ie}{{\it i.e.},\,}
\newcommand{\etal}{{\it et al.\ }}
\newcommand{\eg}{{\it e.g.},\,}
\newcommand{\cf}{{\it cf.\ }}
\newcommand{\vs}{{\it vs.\ }}
\newcommand{\zdot}{\makebox[0pt][l]{.}}
\newcommand{\up}[1]{\ifmmode^{\rm #1}\else$^{\rm #1}$\fi}
\newcommand{\dn}[1]{\ifmmode_{\rm #1}\else$_{\rm #1}$\fi}
\newcommand{\upd}{\up{d}}
\newcommand{\uph}{\up{h}}
\newcommand{\upm}{\up{m}}
\newcommand{\ups}{\up{s}}
\newcommand{\arcd}{\ifmmode^{\circ}\else$^{\circ}$\fi}
\newcommand{\arcm}{\ifmmode{'}\else$'$\fi}
\newcommand{\arcs}{\ifmmode{''}\else$''$\fi}
\newcommand{\MS}{{\rm M}\ifmmode_{\odot}\else$_{\odot}$\fi}
\newcommand{\RS}{{\rm R}\ifmmode_{\odot}\else$_{\odot}$\fi}
\newcommand{\LS}{{\rm L}\ifmmode_{\odot}\else$_{\odot}$\fi}

\newcommand{\Abstract}[2]{{\footnotesize\begin{center}ABSTRACT\end{center}
\vspace{1mm}\par#1\par
\noindent
{~}{\it #2}}}

\newcommand{\TabCap}[2]{\begin{center}\parbox[t]{#1}{\begin{center}
  \small {\spaceskip 2pt plus 1pt minus 1pt T a b l e}
  \refstepcounter{table}\thetable \\[2mm]
  \footnotesize #2 \end{center}}\end{center}}

\newcommand{\TableSep}[2]{\begin{table}[p]\vspace{#1}
\TabCap{#2}\end{table}}

\newcommand{\FigCap}[1]{\footnotesize\par\noindent Fig.\  %
  \refstepcounter{figure}\thefigure. #1\par}

\newcommand{\TableFont}{\footnotesize}
\newcommand{\TableFontIt}{\ttit}
\newcommand{\SetTableFont}[1]{\renewcommand{\TableFont}{#1}}

\newcommand{\MakeTable}[4]{\begin{table}[htb]\TabCap{#2}{#3}
  \begin{center} \TableFont \begin{tabular}{#1} #4 
  \end{tabular}\end{center}\end{table}}

\newcommand{\MakeTableSep}[4]{\begin{table}[p]\TabCap{#2}{#3}
  \begin{center} \TableFont \begin{tabular}{#1} #4 
  \end{tabular}\end{center}\end{table}}

\newenvironment{references}%
{
\footnotesize \frenchspacing
\renewcommand{\thesection}{}
\renewcommand{\in}{{\rm in }}
\renewcommand{\AA}{Astron.\ Astrophys.}
\newcommand{\AAS}{Astron.~Astrophys.~Suppl.~Ser.}
\newcommand{\ApJ}{Astrophys.\ J.}
\newcommand{\ApJS}{Astrophys.\ J.~Suppl.~Ser.}
\newcommand{\ApJL}{Astrophys.\ J.~Letters}
\newcommand{\AJ}{Astron.\ J.}
\newcommand{\IBVS}{IBVS}
\newcommand{\PASP}{P.A.S.P.}
\newcommand{\Acta}{Acta Astron.}
\newcommand{\MNRAS}{MNRAS}
\renewcommand{\and}{{\rm and }}
\section{{\rm REFERENCES}}
\sloppy \hyphenpenalty10000
\begin{list}{}{\leftmargin1cm\listparindent-1cm
\itemindent\listparindent\parsep0pt\itemsep0pt}}%
{\end{list}\vspace{2mm}}

\def\TYLDA{~}
\newlength{\DW}
\settowidth{\DW}{0}
\newcommand{\dw}{\hspace{\DW}}

\newcommand{\refitem}[5]{\item[]{#1} #2%
\def\REFARG{#3}\ifx\REFARG\TYLDA\else, {\it#3}\fi
\def\REFARG{#4}\ifx\REFARG\TYLDA\else, {\bf#4}\fi
\def\REFARG{#5}\ifx\REFARG\TYLDA\else, {#5}\fi.}

\newcommand{\Section}[1]{\section{\hskip-6mm.\hskip3mm#1}}
\newcommand{\Subsection}[1]{\subsection{#1}}
\newcommand{\Acknow}[1]{\par\vspace{5mm}{\bf Acknowledgements.} #1}
\pagestyle{myheadings}

\newfont{\bb}{ptmbi8t at 12pt}
\newcommand{\xrule}{\rule{0pt}{2.5ex}}
\newcommand{\xxrule}{\rule[-1.8ex]{0pt}{4.5ex}}
\def\thefootnote{\fnsymbol{footnote}}
\begin{center}
{\Large\bf The Optical Gravitational Lensing Experiment.
\vskip1pt
Additional Planetary and Low-Luminosity \\
Object Transits from the OGLE 2001 and 2002 \\
Observational Campaigns\footnote{Based on observations
obtained with the 1.3~m Warsaw telescope at the Las Campanas Observatory
of the Carnegie Institution of Washington.}}
\vskip.6cm
{\bf
A.~~U~d~a~l~s~k~i$^1$,~~G.~~P~i~e~t~r~z~y~\'n~s~k~i$^{2,1}$,
~~M.~~S~z~y~m~a~{\'n}~s~k~i$^1$,
~~M.~~K~u~b~i~a~k$^1$,~~K.~~\.Z~e~b~r~u~\'n$^1$,
~~I.~~S~o~s~z~y~\'n~s~k~i$^1$,~~O.~~S~z~e~w~c~z~y~k$^1$,
~~and~~\L.~~W~y~r~z~y~k~o~w~s~k~i$^1$}
\vskip2mm
$^1$Warsaw University Observatory, Al.~Ujazdowskie~4, 00-478~Warszawa, Poland\\
e-mail: (udalski,pietrzyn,msz,mk,zebrun,soszynsk,szewczyk,wyrzykow)@astrouw.edu.pl\\
$^2$ Universidad de Concepci{\'o}n, Departamento de Fisica,
Casilla 160--C, Concepci{\'o}n, Chile
\end{center}

\Abstract{The photometric data collected by OGLE-III during the 2001 and 2002 
observational campaigns aiming at detection of planetary or low-luminosity 
object transits were corrected for small scale systematic effects using the 
data pipeline by Kruszewski and Semeniuk and searched again for low amplitude 
transits. Sixteen new objects with small transiting companions, additional to 
previously found samples, were discovered. Most of them are small amplitude 
cases which remained undetected in the original data. 

Several new objects seem to be very promising candidates for systems 
containing substellar objects: extrasolar planets or brown dwarfs. Those 
include OGLE-TR-122, OGLE-TR-125, OGLE-TR-130, OGLE-TR-131 and a few others. 
Those objects are particularly worth spectroscopic follow-up observations for 
radial velocity measurements and mass determination. With well known 
photometric orbit only a few RV measurements should allow to confirm their 
actual status. 

All photometric data of presented objects are available to the astronomical 
community from the OGLE {\sc Internet} archive.}

\Section{Introduction}
Extensive photometric campaigns conducted by the OGLE-III photometric survey 
in 2001 and 2002 (Udalski \etal 2002abc) proved that the transit photometric 
technique of detection of extrasolar planets can be successfully applied. The 
technique allows to identify an extrasolar planet or other small size object 
when the inclination of its orbit is close to 90\arcd and the small body 
passes (transits) the disk of its host star  every orbital revolution 
obscuring tiny part of its light. The drop of brightness during the 
transit -- of the order of a few percent or less -- and duration of  
transit provide information on the size of small companion. Unfortunately, the 
photometry alone does not allow to unambiguously distinguish between 
extrasolar planets and other astrophysical low-luminosity objects like brown 
dwarfs or very red M-type dwarfs, as all these bodies may have similar 
dimensions. 

On the other hand when the transit method is combined with high accuracy 
radial velocity measurements the degeneracy can be removed as the mass of the 
above objects differs by two orders of magnitude or more. Moreover, in that 
case all the most important parameters of the planet or low-luminosity object 
like its mass, radius and  density can be accurately derived because the 
inclination of orbit is known in the case when transits occur. Thus, the large 
scale photometric survey of large number of stars can potentially provide many 
candidates for extrasolar planetary systems. Then they must be 
spectroscopically followed up for the final confirmation with radial velocity 
techniques. It is worth noticing that because the photometric orbit of the 
candidates is well known the spectroscopic follow up requires much less 
observing time. Also, with the transit method the extrasolar planetary systems 
can be detected at much larger distances than with other methods because 
precise photometry may be obtained for relatively faint stars. 

The search for planetary and low luminosity object transits became one of the 
top priority goals of the OGLE project after the beginning of the third phase 
-- OGLE-III. During the first photometric campaign carried out in June/July 
2001 three fields in the direction of the Galactic center were regularly 
observed. Results of the campaign were presented in Udalski \etal (2002a) and 
in a supplement (Udalski \etal 2002b), after reanalyzing the data with more 
efficient transit detection algorithm of Kov{\'a}cs, Zucker and Mazeh (2002). 
In total 59 objects (two turned out to be the same star observed in 
overlapping region of two fields) revealing small amplitude transits were 
discovered among about 52\,000 disk stars from the nearby Galactic arm. For the 
vast majority of candidates more than two transits were observed allowing 
determination of photometric orbits. Based on the estimation of size and 
shape of the light curve (ellipsoidal variation) it was clear from the very 
beginning that most of the objects found may have rather stellar companions. 
However, in several cases all photometric characteristics indicated very good 
candidates for extrasolar planets or substellar companions. 

The second OGLE-III transit campaign was conducted in February/May 2002. This 
time three fields located in the Carina part of the Galactic disk were 
monitored. Sixty two new objects with transiting companions were discovered in 
the Carina campaign data (Udalski \etal 2002c). Again, in several cases 
photometric data indicated substellar companions. 

In the meanwhile, many groups followed up spectroscopically OGLE-III 
candidates. In January 2003 the first object, namely OGLE-TR-56, was confirmed 
by Konacki \etal (2003) as the first extrasolar planetary system discovered 
with the transit method. A few months later another object, OGLE-TR-3, was 
claimed by Dreizler \etal (2003) to be also a planetary system. 

Detection of transits caused by planetary objects requires extremely high 
accuracy of photometric measurements. For instance, Jupiter would cause a 
transit only about 1\% deep when observed from outside the solar system. This 
is one of the main reasons why the transit method has not been successfully 
applied until now. The smaller the transit depth, the bigger the chance that 
it is caused by a substellar object. 

The OGLE-III photometric data are of high quality due to stable hardware
and a new method of data reduction based on image subtraction techniques
(Alard   2000, Wo{\'z}niak 2000). For the brightest stars the standard
deviation of  the entire data set of 800--1100 observations is usually
smaller than 5  milimagnitudes. Nevertheless, even visual inspection of
the data indicates  that some low level systematic effects can be
present in the OGLE-III data.  For instance, changing zenith angle of
observed fields may produce systematic  errors in photometry at
milimagnitude levels due to differential refraction.  As all kinds of
systematic errors smear out the details in the folded light  curves, it
became clear that non-negligible number of, in particular low 
amplitude, transits could have been lost in the previous analyzes of the
OGLE  data. 

Recently Kruszewski and Semeniuk (2003, in preparation) analyzed the OGLE-III 
data for systematic effects. They prepared a data pipeline that learns about 
the magnitude of the most significant systematic errors based on several 
hundreds best photometry stars and then corrects photometry of all stars for 
those systematic effects. 

In this paper we present results of the repeated search for transits in the 
OGLE-III photometric data from 2001 and 2002 campaigns corrected for 
systematic effects with the Kruszewski and Semeniuk (2003, in preparation) 
data pipeline. As we expected, we discovered additional 16 objects with 
transiting companions lost in the noise in our previous analyzes. Several of 
our new candidates reveal transits of very small amplitude indicating 
Jupiter-size companions. They are very good candidates for  transiting 
extrasolar planets. Similarly to our previous transit samples the photometric 
data of our new candidates are available to the astronomical community from 
the OGLE {\sc Internet} archive. 

\Section{Observational Data}
All observations presented in this paper were collected with the 1.3-m Warsaw  
telescope at the Las Campanas Observatory, Chile (operated by the Carnegie  
Institution of Washington), during the third phase of the OGLE project. The 
telescope was equipped with a wide field CCD mosaic camera consisting of eight 
${2048\times4096}$ pixel SITe ST002A detectors. The pixel size of each 
detector is 15~$\mu$m giving the 0.26~arcsec/pixel scale at the focus of the 
Warsaw telescope. Full field of view of the camera is about 
${35\arcm\times35\arcm}$. The gain of each chip is adjusted to be about  
1.3~e$^-$/ADU with the readout noise of about 6 to 9~e$^-$, depending on chip. 
 
The 2001 OGLE transit campaign lasted from June 12 to July 28, 2001. More than 
800 epochs of three fields in the direction of the Galactic center were 
collected on 32 nights. The exposure time was set to 120 seconds, and all 
fields were observed every 12 minutes. The photometric data of the 2002 
campaign were collected during 76 nights spanning 95 days starting from 
February 17, 2002. Three fields located in the Carina region of the Galactic  
disk were observed continuously with the time resolution of about 15 minutes. 
In total, more than 1100 epoch were collected for each field. The exposure 
time was 180 second in that case. All observations were collected with the 
{\it I}-band filter. For more details on the observing strategy, observed 
fields and photometry techniques the reader is referred to the original papers 
on the 2001 (Udalski \etal 2002a) and 2002 campaigns (Udalski \etal 2002c). 

The photometric data from the original transit campaigns were additionally 
supplemented with  random epoch photometric observations collected to the end 
of May 2003. The number of additional observations reached 300 in the case of 
the 2001 campaign data and several or several tens in the case of 2002 
campaign data. 

\Section{Correction for Systematic Effects and Transit Search}
All stars with accurate enough photometry (standard deviation of all 
observations not larger than 0.015~mag) that were used for transit search in 
the previous papers (about 52\,000 and 103\,000 for 2001 and 2002 campaigns, 
respectively) were subject to the procedure of removing effects of small 
systematic errors in the data. The data pipeline was written by Kruszewski and 
Semeniuk (2003, in preparation) and will be described in detail in a separate 
paper. In short, the software corrects the data for effects of differential 
refraction, and small drifts of the magnitude scale and its zero point. Based 
on several hundreds best constant stars selected for each field, it calculates  
the magnitude of the systematic effects. Then all stars from the sample are 
corrected for these effects. 

In the next step, corrected photometry of all stars was searched for transits 
similar to the original papers. We again used the fast and efficient BLS 
algorithm  of Kov{\'a}cs, Zucker and Mazeh (2002). We used the same parameters 
of the BLS algorithm as in the 2001 and 2002 campaigns (Udalski \etal 2002b)  
and limited our search for transits to periods from 1.05 to 10~days. 

In the final step the light curves of all candidates were inspected visually. 
We removed from our list of transit candidates all object already presented in 
Udalski \etal (2002abc) and suspected objects found during earlier analyzes. 
We discovered many new objects, usually with very small amplitudes which were 
not detected previously. However, on the final list of additional transit 
objects only  those that have a significant probability of being a true 
transit were left. Therefore we removed a large number of new detections that 
were evidently grazing eclipses (small amplitude V-shape of eclipses) or clear 
blends of an eclipsing system with a star. 

\Section{Discussion}
Sixteen objects with transiting planetary or low-luminosity companions 
remained on our list of new candidates found in the 2001 and 2002 OGLE-III 
photometry. Table~1 presents all basic data on these objects. The notation of 
objects follows that used by OGLE  for previous transit campaigns. Therefore 
the first object in Table~1 is designated as OGLE-TR-122. 
\MakeTable{l@{\hspace{7pt}}
c@{\hspace{7pt}}c@{\hspace{7pt}}c@{\hspace{7pt}}r@{\hspace{7pt}}
c@{\hspace{7pt}}c@{\hspace{7pt}}r}{12.5cm}
{New OGLE-III planetary and low luminosity object transits} 
{\hline
\noalign{\vskip4pt} 
\multicolumn{1}{c}{Name} & RA (J2000)  & DEC (J2000) &   $P$    &
\multicolumn{1}{c}{$T_0$}& $I$    &$\Delta I$  &  \multicolumn{1}{c}{$N_{\rm tr}$}\\
           &             &             & [days]   & --2452000     &[mag] &[mag] \\ 
\noalign{\vskip4pt}
\hline
\noalign{\vskip4pt}
OGLE-TR-122 & 11\uph06\upm51\zdot\ups99 & $-60\arcd51\arcm45\zdot\arcs7$ & 7.26867 & 342.28258 & 15.61  & 0.019 &  5 \\
OGLE-TR-123 & 11\uph06\upm51\zdot\ups19 & $-61\arcd11\arcm10\zdot\arcs1$ & 1.80380 & 324.97937 & 15.40  & 0.008 & 12 \\
OGLE-TR-124 & 10\uph59\upm49\zdot\ups57 & $-61\arcd52\arcm07\zdot\arcs5$ & 2.75330 & 327.35782 & 15.10  & 0.011 &  5 \\
OGLE-TR-125 & 10\uph57\upm51\zdot\ups95 & $-61\arcd43\arcm58\zdot\arcs3$ & 5.30382 & 343.82550 & 15.82  & 0.013 &  4 \\
OGLE-TR-126 & 10\uph59\upm42\zdot\ups01 & $-61\arcd32\arcm34\zdot\arcs2$ & 5.11080 & 327.40830 & 15.57  & 0.022 &  3 \\
OGLE-TR-127 & 10\uph55\upm10\zdot\ups25 & $-61\arcd31\arcm04\zdot\arcs9$ & 1.92720 & 329.84698 & 14.97  & 0.011 &  8 \\
OGLE-TR-128 & 10\uph54\upm27\zdot\ups54 & $-61\arcd38\arcm17\zdot\arcs3$ & 7.39100 & 327.42530 & 15.00  & 0.016 &  5 \\
OGLE-TR-129 & 10\uph50\upm51\zdot\ups12 & $-61\arcd34\arcm55\zdot\arcs2$ & 5.74073 & 327.36759 & 16.19  & 0.034 &  3 \\
OGLE-TR-130 & 10\uph51\upm23\zdot\ups54 & $-61\arcd47\arcm22\zdot\arcs0$ & 4.83037 & 327.28057 & 15.94  & 0.028 &  5 \\
OGLE-TR-131 & 10\uph50\upm04\zdot\ups86 & $-61\arcd51\arcm32\zdot\arcs5$ & 1.86990 & 324.94513 & 15.69  & 0.011 &  9 \\
OGLE-TR-132 & 10\uph50\upm34\zdot\ups72 & $-61\arcd57\arcm25\zdot\arcs9$ & 1.68965 & 324.70067 & 15.72  & 0.011 & 11 \\
OGLE-TR-133 & 17\uph49\upm51\zdot\ups32 & $-29\arcd56\arcm20\zdot\arcs4$ & 5.31075 &  83.34913 & 16.63  & 0.034 &  3 \\
OGLE-TR-134 & 17\uph52\upm29\zdot\ups95 & $-29\arcd33\arcm01\zdot\arcs7$ & 4.53720 &  79.91817 & 13.49  & 0.011 &  2 \\
OGLE-TR-135 & 17\uph52\upm24\zdot\ups27 & $-29\arcd39\arcm22\zdot\arcs4$ & 2.57330 &  79.83362 & 15.16  & 0.019 &  5 \\
OGLE-TR-136 & 17\uph54\upm59\zdot\ups23 & $-29\arcd19\arcm39\zdot\arcs1$ & 3.11580 &  76.10417 & 14.93  & 0.016 &  7 \\
OGLE-TR-137 & 17\uph56\upm12\zdot\ups65 & $-29\arcd45\arcm03\zdot\arcs1$ & 2.53782 &  75.64246 & 15.85  & 0.030 &  7 \\
\hline}

In the subsequent columns of Table~1 the following data are provided: 
Identification, equatorial  coordinates (J2000), orbital period, epoch of 
mid-eclipse, {\it I}-band magnitude outside transit, the depth of transit, and 
number of transits observed ($N_{\rm tr}$). Accuracy of the magnitude scale is 
about 0.1--0.2~mag. In Appendix the light curves with close-ups around the 
transit and finding charts of all objects from Table~1 are presented. The 
finding chart is a ${60\arcs\times60\arcs}$ subframe of the {\it I}-band 
reference image centered on the star. The star is marked by a white cross. 
North is up and East to the left in these images. 

The observed transits can be caused by extrasolar planets or brown dwarfs or 
small late  M-type dwarfs. Radial velocity measurements are necessary to 
distinguish between these possibilities. It is worth noticing that the new 
objects usually show small transit depth. Therefore, the new sample includes 
several objects that belong to the best OGLE-III candidates for extrasolar 
planetary systems.
\renewcommand{\TableFont}{\footnotesize}
\MakeTable{lccccc}{12.5cm}{Dimensions of stars and companions for central
passage ($i=90\arcd$) \\ and amplitudes of $\cos P$ and $\cos 2P$ variability.}
{\hline
\noalign{\vskip3pt}
Name       & $R_s$   & $R_c$ & $M_s$ & $a_{c1}$ & $a_{c2}$\\
&[\RS]&[\RS]&[\MS]&mmag&mmag\\
\noalign{\vskip3pt}
\hline
\noalign{\vskip3pt}
OGLE-TR-122 &  0.72 & 0.086  & 0.66 & $0.43 (-2.31 \pm 1)$ & $0.40 (-0.84 \pm 1)$ \\
OGLE-TR-123 &  2.56 & 0.205  & 3.24 & $0.40 (+0.42 \pm 1)$ & $0.39 (+4.52 \pm 1)$ \\
OGLE-TR-124 &  0.52 & 0.047  & 0.44 & $0.29 (+0.39 \pm 1)$ & $0.27 (+0.30 \pm 1)$ \\
OGLE-TR-125 &  1.42 & 0.142  & 1.56 & $0.50 (-0.11 \pm 1)$ & $0.47 (+0.31 \pm 1)$ \\
OGLE-TR-126 &  1.06 & 0.138  & 1.08 & $0.35 (-1.56 \pm 1)$ & $0.33 (+1.07 \pm 1)$ \\
OGLE-TR-127 &  0.74 & 0.066  & 0.68 & $0.34 (-0.07 \pm 1)$ & $0.33 (+0.45 \pm 1)$ \\
OGLE-TR-128 &  5.06 & 0.556  & 7.58 & $0.51 (+0.42 \pm 1)$ & $0.48 (+2.97 \pm 1)$ \\
OGLE-TR-129 &  0.68 & 0.108  & 0.61 & $0.57 (-0.78 \pm 1)$ & $0.49 (+0.78 \pm 1)$ \\
OGLE-TR-130 &  0.30 & 0.042  & 0.22 & $0.43 (+0.83 \pm 1)$ & $0.40 (-0.61 \pm 1)$ \\
OGLE-TR-131 &  0.65 & 0.059  & 0.59 & $0.43 (+1.11 \pm 1)$ & $0.41 (+2.02 \pm 1)$ \\
OGLE-TR-132 &  1.05 & 0.094  & 1.06 & $0.34 (+2.26 \pm 1)$ & $0.33 (+1.49 \pm 1)$ \\
OGLE-TR-133 &  0.50 & 0.081  & 0.42 & $0.66 (+1.52 \pm 1)$ & $0.67 (-0.95 \pm 1)$ \\
OGLE-TR-134 &  1.15 & 0.103  & 1.19 & $0.58 (-0.42 \pm 1)$ & $0.58 (-0.41 \pm 1)$ \\
OGLE-TR-135 &  2.72 & 0.327  & 3.50 & $0.67 (+1.73 \pm 1)$ & $0.63 (+5.43 \pm 1)$ \\
OGLE-TR-136 &  1.88 & 0.207  & 2.20 & $0.60 (+0.87 \pm 1)$ & $0.54 (+1.10 \pm 1)$ \\
OGLE-TR-137 &  1.05 & 0.157  & 1.06 & $0.74 (+1.73 \pm 1)$ & $0.66 (+2.34 \pm 1)$ \\
\hline}

However, one should remember that blending of a regular eclipsing star with 
total eclipses and  a close optical or physically related (wide binary system)  
unresolved in the seeing disk neighbor can produce a light curve mimicking 
transits. Blends with separation larger than about 0\zdot\arcs4 can be 
practically ruled out in our samples as we always verify coincidence of the 
centroid of star in the reference image with the centroid of the loss of light 
in the subtracted image (taken during transit). When the separation is larger 
than that limit the object is removed from our list. Nevertheless, closer 
blends cannot be excluded. Therefore some of our new candidates can actually 
be faked transits caused by blending effect. Probability of blending is much 
higher in the 2001 campaign data because of strong stellar background by the 
Galactic bulge stars. The blending problem should also be clarified by high 
resolution spectroscopy. 

In Table~2 we present estimation of the size of transiting objects in our new 
candidates and host stars calculated in similar way to the previous papers 
(Udalski \etal 2002abc). Unfortunately, without any additional information on 
the radius of the primary it is not possible to obtain actual size of the 
companion when the errors of individual observations are comparable to the  
transit depth. Due to well known degeneracy between radii of the host star and  
companion, $R_s$ and $R_c$, inclination $i$, and limb darkening $u$, similar  
quality photometric solutions can be obtained for different inclinations of  
the orbit and radii of components (in the {\it I}-band the transit light curve  
is practically insensitive to the limb darkening parameter $u$) making 
selection of the proper solution practically impossible. 

Because the radius of the primary cannot be presently constrained as we do not 
know spectral types of primary stars, and their colors can be affected to 
unknown degree by the interstellar extinction only the lower limit on the size 
of the companion can be calculated assuming that the transit is central, \ie  
${i=90\arcd}$. The corresponding radius of the primary is also the lower 
limit. In Table~2 we list dimensions of the primary and low-luminosity 
companion assuming additionally that the host star follows the mass-radius 
relation for main sequence stars ($R/\RS=(M/\MS)^{0.8}$). The mass of the 
primary is also listed in Table~2. In practice the transits might be 
non-central, \ie the size of the star and companion can be larger than given 
in Table~2. Also when the host star is evolved the estimations can be 
inaccurate. Solid line in the close-up windows in Appendix shows the transit 
model light curve calculated for the central passage. In some cases the fit 
is not satisfactory indicating the inclination smaller than $90\arcd$. 
However, in most cases the central passage fit is practically 
indistinguishable from others so at this stage it is impossible to derive 
other values than the lower limits of radii provided in Table~2. 

In the last two columns of Table~2 we list the amplitudes and errors of the 
periodic sinusoidal variation with period $P$ ($a_{c1}$) and $2P$ ($a_{c2}$) 
fitted to the photometric data. Periodicity $P$ is interpreted as an 
reflection effect while $2P$ as an ellipsoidal effect. Presence of the latter 
at statistically significant level immediately indicates stellar companion or 
a blend of eclipsing system with nearby star (Drake 2003, Sirko and 
Paczy{\'n}ski 2003). The amplitudes are provided in the same form as in Sirko 
and Paczy{\'n}ski (2003): $b_i(a_{ci}/b_i \pm 1)$ where $a_{ci}$ is the 
amplitude and $b_i$ its error. In this way one can immediately see the ratio 
of amplitude to its error, \ie significance of the effect. 

Transits in a few newly discovered objects are certainly caused by stellar 
companions. The light curve of such systems exhibits clearly a statistically 
significant ellipsoidal effect (\eg OGLE-TR-123, OGLE-TR-128, OGLE-TR-135) 
what indicates that the companion mass is relatively large. In the case of 
OGLE-TR-128 perhaps even hints of a secondary eclipse are seen in the light 
curve. Also our limits on size of companions from Table~2 suggest evidently 
stellar companions in all these cases. The remaining objects do not show 
statistically significant ellipsoidal variation or reflection effect. 

On the other hand, there is a group of several objects where the transiting 
companions can be much smaller and those stars are very good candidates for 
extrasolar planetary systems. In particular, the objects from the Carina 
region of the Galactic disk, namely OGLE-TR-122, OGLE-TR-125, OGLE-TR-127, 
OGLE-TR-129, OGLE-TR-130, or OGLE-TR-131 belong to the most promising, from 
the photometric point of view, candidates for planetary systems. This group 
of stars is particularly worth spectroscopic follow up observations as the 
chance that the companions are substellar objects is larger than in other 
cases. It is worth noting that additional transits of OGLE-TR-122 and  
OGLE-TR-125 were observed in 2003 season. Therefore the photometric 
ephemerides of these two objects are much more accurate than of others. The 
new detections from 2001 campaign from the Galactic center are less promising 
for extrasolar planet discoveries. Only OGLE-TR-133 and OGLE-TR-134 and perhaps 
OGLE-TR-137 may host substellar companions. The latter object might, however, 
display a weak ellipsoidal variation. 

Results of our new search for transits in the corrected OGLE-III data indicate 
that careful analysis of the systematic errors significantly improves quality 
of photometry. Sixteen new detections increase the number of OGLE objects with 
transiting companions by 13\%. However, most of the new objects reveal low 
depth transits and therefore are very good candidates for extrasolar planetary 
systems. Therefore,  we plan to run the Kruszewski and Semeniuk (2003, in 
preparation) algorithm also on our other photometric data, namely collected 
during 2003 transit campaigns. 

The photometric data of all new objects with transiting companions discovered  
in the OGLE-III 2001 and 2002 photometric data corrected for systematic 
effects  are available in the electronic form from  the OGLE archive: 
\vskip3pt
\centerline{\it http://ogle.astrouw.edu.pl} 
\vskip3pt
\centerline{\it ftp://ftp.astrouw.edu.pl/ogle/ogle3/transits/new\_2001\_2002}
\vskip3pt
\noindent
or its US mirror
\vskip3pt
\centerline{\it http://bulge.princeton.edu/\~{}ogle}
\vskip3pt
\centerline{\it ftp://bulge.princeton.edu/ogle/ogle3/transits/new\_2001\_2002}

\Acknow{We would like to thank Prof.\ A.~Kruszewski and Dr.\ I.~Semeniuk for 
making the data pipeline available for our project before publication. We are 
very grateful to Mr.\ E.~Sirko for calculation of amplitudes of the reflection 
and ellipsoidal variation of our objects. We also thank Prof.\ B.\ 
Paczy{\'n}ski for many interesting comments. The paper was partly supported by 
the  Polish KBN  grant 2P03D02124 to A.\ Udalski. Partial support to the OGLE  
project was provided with the NSF grant AST-0204908 and NASA grant NAG5-12212 
to B.~Paczy\'nski. A.~Udalski acknowledges support from the grant ``Subsydium 
Profesorskie'' of the Foundation for Polish Science.}

\end{document}